\documentclass[12pt]{article}
\usepackage{scicite}
\usepackage{graphicx}
\usepackage{times}
\usepackage{moreverb,url}
\usepackage{multirow}
\usepackage[colorlinks,bookmarksopen,bookmarksnumbered,citecolor=red,urlcolor=red]{hyperref}
\newcommand\BibTeX{{\rmfamily B\kern-.05em \textsc{i\kern-.025em b}\kern-.08em
T\kern-.1667em\lower.7ex\hbox{E}\kern-.125emX}}
\newenvironment{sciabstract}{%
\begin{quote} \bf}
{\end{quote}}

\begin{document}

%\runninghead{Mori\~na, Puig and Navarro}

\title{Analysis of zero inflated dichotomous variables from a Bayesian perspective: Application to occupational health}

\author
{David Moriña$^{1,2\ast}$, Pedro Puig$^{2,3}$ and Albert Navarro$^{4,5}$\\
\\
\normalsize{$^{1}$Department of Econometrics, Statistics and Applied Economics,}\\
\normalsize{Riskcenter-IREA, University of Barcelona (UB)}\\
\normalsize{$^{2}$Centre de Recerca Matem\`atica (CRM)}\\
\normalsize{$^{3}$Barcelona Graduate School of Mathematics (BGSMath),}\\
\normalsize{Department of Mathematics, Autonomous University of Barcelona (UAB)}\\
\normalsize{$^{4}$Research group on Psychosocial Risks, Organization of Work and Health (POWAH),}\\
\normalsize{Autonomous University of Barcelona (UAB)}\\
\normalsize{$^{5}$Biostatistics Unit, Faculty of Medicine, Autonomous University}\\
\normalsize{of Barcelona}\\
\\
\\
\normalsize{$^\ast$To whom correspondence should be addressed; E-mail:  dmorina@ub.edu	}
}

% Include the date command, but leave its argument blank.

\date{}

%\keywords{presenteeism, bayesian methods, zero-inflation, simulation study, Bernoulli mixture models}

\maketitle

\begin{sciabstract}
This work proposes a new methodology to fit zero inflated Bernoulli data from a Bayesian approach, able to distinguish between two potential sources of zeros (structurals and non-structurals). Its usage is illustrated by means of a real example from the field of occupational health as the phenomenon of sickness presenteeism, in which it is reasonable to think that some individuals will never be at risk of suffering it because they have not been sick in the period of study (structural zeros). Without separating structural and non-structural zeros one would one would be studying jointly the general health status and the presenteeism itself, and therefore obtaining potentially biased estimates as the phenomenon is being implicitly underestimated by diluting it into the general health status. The proposed methodology performance has been evaluated through a comprehensive simulation study, and it has been compiled as an R package freely available to the community.
\end{sciabstract}

\section{Introduction}\label{intro}
In general, zero-inflated models are aimed to addressing the problem that arises from having two different sources that generate the zero values observed in a distribution. In practice, this is due to the fact that the population studied actually consists of two subpopulations: one in which the value zero is by default (structural zero) and the other is circumstantial (sample zero). An example could be the study of sickness presenteeism (SP), i.e. attending work while sick \cite{Johns2010}. If it is not previously restricted, the population is made up, among others, of workers who are zero because they have never been sick (structural zeros) and workers who, having been sick, did not attend their work place (sample zeros). Note that the difference is important: roughly the first zero informs us exclusively about the status of health, the second about the exercise of the right to take a sick leave.

The most used zero-inflated models are those that are related to counting variables, where it is considered that the zero value has a dichotomous source that determines whether or not the study is at risk and another source, corresponding to the individuals at risk , which models the number of episodes (counts) that have been presented. In this context, the available models would be Zero-Inflated Poisson (ZIP) and Negative Binomial (ZINB). The zero-inflated models considering that both sources are dichotomous (a mixture of two Bernoulli random variables, one with probability of success
$\omega$ and the other with probability of success $p$) are much less studied in practice. This is due, in large part, to the fact that the resulting distribution is once again a Bernoulli with probability of success $\omega \cdot p$, so that the proportion of structural zeros $(1 - \omega)$ and sample zeros $(1- p)$ are indistinguishable from the point of view of frequentist statistics. However, from the Bayesian perspective and using known reasonable information about these proportions, it is possible to distinguish the two sources of zeros and estimate $\omega$ and $p$.

Some authors have recently suggested, in other areas such as the classification or identification of images, the usage of Bernoulli-mixture models, based on numerical algorithms such as Expectation-Maximization (EM) to estimate the parameters \cite{Barbu2015, Diop2016}, given the complexity of the likelihood functions involved. In these cases, however, the inclusion of covariates or adjustment variables is virtually impossible. Also in other areas there are some recent developments in a similar line, such as \cite{Sun2007}.

In this article we illustrate the use of Zero Inflated Bernoulli (ZIB) models by means of a real dataset on SP, and the results obtained are compared with those of adjusted logistic regressions on the total population or only in those individuals at risk. In the literature, the SP registry is carried out in a self-reported way, asking about the episodes in the last year and later, recorded in a dichotomized way (no SP: 0 episodes; yes SP: 1 or more episodes). The justification for this dichotomization is fundamentally based on two aspects: one, the possible memory bias; second, the excessive influence of workers who report a very high number of episodes.

\section{Methods}\label{methods}
\begin{figure}[h]
  \caption{\label{model_schema}Model schema. Only variable Y is observed.}
  \centering
    \includegraphics[width=0.35\textwidth]{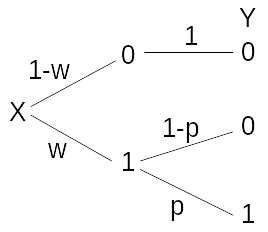}
\end{figure}

Let $Y$ be the variable that indicates occurrence of the phenomenon under study. The proposed model has a probability function defined by
\begin{equation}\begin{array}{ll}
 P(Y=0) & = (1-\omega) + \omega \cdot (1-p) \\
 P(Y=1) & = \omega \cdot p, \\
\end{array}\end{equation}
where $\omega$ is the probability of exposure and $p$ is the probability of occurrence of the phenomenon of interest among exposed individuals, as shown in Figure~\ref{model_schema}. According to this scheme, the proportion of structural zeros will be $1-\omega$ and the proportion of non-structural zeros will be
$\omega \cdot (1-p)$.

To overcome the impossibility of models without covariates based on the frequentist approach to differentiate between structural and non-structural zeros, in this work a model within the Bayesian framework is proposed. In this context, we assume that the prior distribution of the parameter of the first Bernoulli $\omega$ is uniform between 0 and 0.5 while the prior distribution of the probability of success in the second Bernoulli $p$ is uniform between 0.5 and 1. Obviously, these hypotheses can be modified according to the a priori knowledge of the parameters that govern the phenomenon to be studied. In this way, the proposed model will be able to distinguish the two sources of zeros. If covariates are included in the model, this distinction is not necessary since the covariates themselves allow the origin of the zeros to be distinguished.
To ensure that the estimates are kept within the appropriate parameter space, the \textit{logit} link, commonly used in logistic regression, has been used.

\begin{equation}\begin{array}{ll}
 logit(\omega) = \log(\frac{\omega}{1-\omega}) &= \displaystyle \theta_0 + \theta_1 X_1 + \ldots + \theta_k X_k \\
 logit(p) = \log(\frac{p}{1-p}) & = \displaystyle \beta_0 + \beta_1 Z_1 + \ldots + \beta_m Z_m,
\end{array}\end{equation}

where $X_1, \ldots, X_k$ are the covariates that have a hypotetical impact over the zero inflated part and $Z_1, \ldots, Z_m$ are the covariates that might have an influence over the non zero inflated part. The parameters $\theta_i$, $i = 0, \ldots, k$ and $\beta_j$, $j = 0, \ldots, m$ are assumed to follow a normal distribution with mean 0 and variance $\sigma^2_{\theta}$ and $\sigma^2_{\beta}$ respectively, modeled as hyperparameters.

The models proposed to analyze the data described in the following section and in the simulation study have been written in the programming language \textit{Stan}, within the \textit{R} environment \cite{cran} and are freely available from the authors as a package called \textit{bayesZIB} \cite{}. The use of the package is very similar to other packages that implement inflated zero models, such as \textit{pscl} \cite{pscl}, to facilitate the interpretation of the results, while a more advanced user could easily adapt the code to their specific requirements.

\subsection{No covariates}
In the particular case in which the interest is in estimating the proportion of structural (1-$\omega$) and sample ($p$) zeros without accounting for the effect of any covariate, the \textit{posterior} distributions of $\omega$ and $p$ can be obtained analytically assuming some \textit{a priori} knowledge of their distributions. As mentioned before, one could set $\omega$ to be uniform distributed on $[0, 0.5]$ and $p$ to be uniform distributed on $[0.5, 1]$. Because the observations are Bernoulli($p \cdot \omega$) distributed, the likelihood function can be written as

\begin{equation}\label{model_likelihood}
L \sim (p \cdot \omega)^m \cdot (1-p \cdot \omega)^{n-m},
\end{equation}

where $m$ is the frequency of occurrence of the phenomenon of interest and $n$ is the total number of observations. From here, the joint \textit{posterior} could be obtained as

\begin{equation}\begin{array}{lll}\label{joint_posterior}
f(p, \omega) & \sim & (p \cdot \omega)^m \cdot (1-p \cdot \omega)^{n-m} \cdot \\ & & U_{[0, 1/2]}(\omega) \cdot U_{[1/2, 1]}(p)
\end{array}\end{equation}

From here the \textit{posterior} marginal distributions of the two parameters can be obtained as

\begin{equation}\begin{array}{lll}\label{posterior_w}
f(\omega) & \sim & \displaystyle \omega^m \int_{1/2}^1 p^m \cdot (1-p \cdot \omega)^{n-m} dp \sim \\
 & & \displaystyle \frac{1}{\omega} \cdot \int_{\omega/2}^{\omega} t^m \cdot (1-t)^{n-m} dt \sim \\
 & & \displaystyle \frac{F(\omega, m+1, n-m+1) - F(\frac{\omega}{2}, m+1, n-m+1)}{\omega} \\
f(p) & \sim & \displaystyle p^m \int_{0}^{1/2} \omega^m \cdot (1-p \cdot \omega)^{n-m} d \omega \sim \\
 & & \displaystyle \frac{1}{p} \cdot \int_{0}^{p/2} t^m \cdot (1-t)^{n-m} dt \sim \\
 & & \displaystyle \frac{F(\frac{p}{2}, m+1, n-m+1)}{p}, 
\end{array}\end{equation}

where $F$ is the beta distribution function with parameters $m+1$ and $n-m+1$, implemented in the \textit{R} function \textit{pbeta}.

\section{Results}\label{results}
This section presents the results of the analyses using the proposed methodology over a real data set and they are compared to the most common alternatives. The performance of the method is also studied by means of a comprehensive simulation study, with and without covariates.

\subsection{Real data}\label{rdata}
In the database used to exemplify the use of the proposed methodology, we have a total of 1564 workers. Among these, it is known that 946 (around 61\%) were not at risk of being presenteeist because they were not ill on any day during the study period. These observations correspond to the concept of structural zeros $(1- \omega = 0.61)$, and an estimate of their proportion can be obtained by using inflated zero models, even taking into account the values of the variables used as explanations in the regression model. The proportion of presenteeists among those exposed is $p = 0.70$. In the following subsections CI is used as an abbreviation of the confidence interval for frequentist analyses or credibility interval when referring to the proposed Bayesian model.

\subsubsection{Including the whole population} 
\begin{figure}[h]
  \caption{\label{priors_posteriors}Prior (left column) and posterior (right column) distributions for $\omega$ and $p$.}
  \centering
    \includegraphics[width=0.8\textwidth]{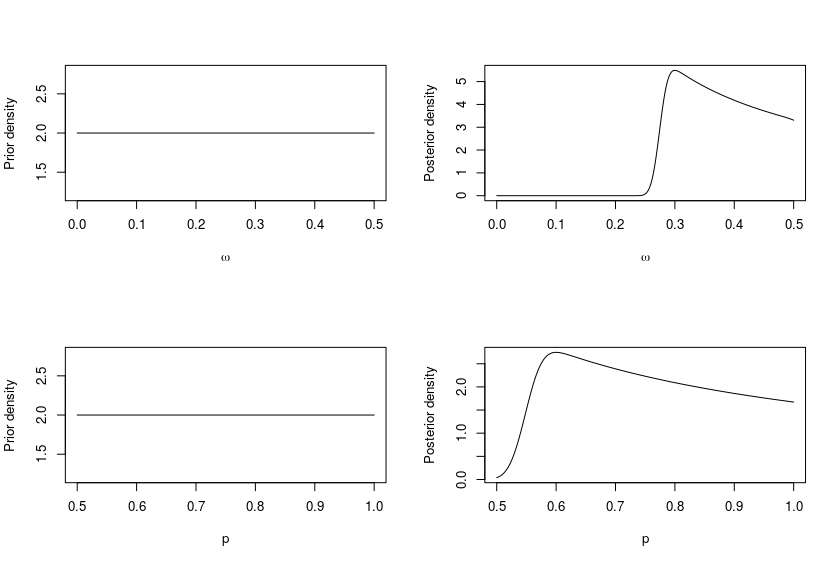}
\end{figure}

Taking all the population into consideration ($n=1564$), i. e., including individuals at risk and not at risk (those who were not at risk during the study period), we fit a Bayesian zero inflated Bernoulli model, where the proportion of structural zeros $1- \omega$ is greater than 0.5 (prior uniform for $\omega$ at $[0, 0.5]$) and the proportion of sample zeros $1-p$ is less than 0.5 (prior uniform for $p$ at $[0.5, 1]$). This information is extracted from \cite{Navarro2018}. In this case, without using covariates, the model allows estimating the values of $\hat{\omega} = 0.37$ (95\% CI: 0.27 - 0.49) and $\hat{p} = 0.74$ (95\% CI : 0.55 - 0.99). Here $\hat{\omega}$ and $\hat{p}$ indicate the median of the marginal posterior of $\omega$ and $p$ respectively. The \textit {a priori} and \textit{a posteriori} marginal distributions of both parameters are shown in Figure~\ref{priors_posteriors}. The different shapes between marginal priors and posteriors show that the models learn from the data.

On the other hand, analyzing these data as is traditional in the literature, using a logistic regression model without taking into account that there are subjects who have not been at risk, the proportion obtained from presentists is 0.27 (95\% CI: 0.25 - 0.30) , a value with a controversial interpretation since it is significantly underestimating the proportion of presenters if the subjects who have not been at risk of being present are excluded, since it is ultimately an estimate of $\hat{\omega} \cdot \hat{p}$, being impossible to identify the two parameters.

Additionally, the proposed model allows incorporating covariates in both processes. To illustrate how it works, we will consider here the self-perceived general state of health (categorized as good or bad) and the feeling of being replaceable, which is an item included in the vulnerability dimension of the Employment Precariousness Scale \cite{Vives2013}, with categories ``Always", ``Sometimes" and ``Never". The hypothesis is that the general state of health would be related to the risk of being present (zero inflated part of the model) and the feeling of being replaceable would be related to presenting the phenomenon once a worker is exposed (non zero inflated part in the model), so the model is including one covariate in each part ($k=m=1$). In this case, as can be seen in the next section, the results of the model shown in Table~\ref{tab1} largely coincide with the standard logistic analyses reported in Table~\ref{tab2}, particularly in the sense and impact of associations. All \textit{R} codes used in this paper are available as Supplementary Material.

\begin{table}[h]
\small\sf\centering
\caption{Bayesian analysis on whole population. CI stands for credible interval\label{tab1}}
\begin{tabular}{ccc}
\hline
 & Covariate & Coefficient. (95\% CI) \\
\hline
\multirow{2}{*}{Struct.} & Intercept & -0.67 (-0.99, 0.20) \\
& Bad & 1.31 (0.78, 3.56) \\
\hline
\multirow{3}{*}{Non-struct.} & Intercept & 1.92 (0.20, 4.31) \\
& Sometimes & 0.09 (-1.42, 3.46) \\
& Never & -1.04 (-2.91, -0.20) \\
\hline
\end{tabular}
\end{table}

Similarly to other regression models, the effect of never having the feeling of being replaceable over the odds of attending work when sick for someone who is at risk compared to workers who always have that sensation can be quantified by $e^{-1.04}=0.35$.

\subsubsection{Excluding healthy population}
If the information about which subjects are actually exposed to the phenomenon of interest is available (an ideal but unusual situation in practice), unexposed subjects could be excluded and a logistic regression model could be adjusted to the exposed individuals. Using the same explanatory variables described in the previous section, the corresponding coefficients and their 95\% confidence intervals are described in Table~\ref{tab2}.

\begin{table}[h]
\small\sf\centering
\caption{Logistic regression on all population and only on exposed individuals. CI stands for confidence interval\label{tab2}}
\begin{tabular}{ccc}
\hline
Population & Covariate & Coefficient. (95\% CI) \\
\hline
\multirow{2}{*}{Whole population} & Intercept & -1.07 (-1.19, -0.95) \\
& Bad & 1.05 (0.68, 1.42) \\
\hline
\multirow{3}{*}{Only exposed} & Intercept & 1.16 (0.79, 1.53) \\
& Sometimes & -0.02 (-0.60, 0.56) \\
& Never & -0.54 (-0.97, -0.11) \\
\hline
\end{tabular}
\end{table}

\subsection{Simulation study}\label{simu}
In order to check the performance of the proposed methodology, 100 random samples were generated for each considered sample size ($n = 500, 1500$), and combination of parameters. The zero inflated part was build upon the logistic regression model $logit(P(X=1)) = \theta_0 + \theta_1 \cdot x_1 + \theta_2 \cdot x_2$, where $x_1$ and $x_2$ are two independent covariates, each following a standard normal distribution. The non zero inflated part was build upon the logistic regression model $logit(P(Y=1 \mid X=1)) = \beta_0 + \beta_1 \cdot x_3 + \beta_2 \cdot x_4$, where $x_3$ and $x_4$ are two independent covariates, each with a standard normal distribution. To cover different effect magnitudes, the following values for each parameter were considered:
\begin{itemize}
 \item $\beta_0 = 0.5, 1, 2$
 \item $\beta_1 = 2, 3, 4$
 \item $\beta_2 = 3$
 \item $\theta_0 = -0.5, -1, -2$
 \item $\theta_1 = -2, -3, -4$
 \item $\theta_2 = -3$
\end{itemize}
For each random sample, the \textit{posterior} marginal distributions of the parameters have been summarised by their median and percentiles 2.5\% and 97.5\%.

Tables~\ref{tab3_1} and~\ref{tab3_2} show, for each combination of parameters, the average estimates and upper and lower limits of the 95\% credibility intervals. As no relevant differences were observed regarding sample sizes, Tables~\ref{tab3_1} and~\ref{tab3_2} shows only the results corresponding to $n=1500$. The results corresponding to $n=500$ are available as Supplementary Material.

\begin{table*}[h!]
\tiny\sf\centering
\caption{Simulation study results including covariates (I).\label{tab3_1}}
\begin{tabular}{p{0.4cm}p{0.4cm}p{0.4cm}p{0.4cm}p{1.50cm}p{1.50cm}p{1.50cm}p{1.50cm}p{1.50cm}p{1.50cm}} % columns
\hline                                          
$\beta_0$ & $\beta_1$ & $\theta_0$ & $\theta_1$ & $\hat{\beta_0}$ (95\% CI) & $\hat{\beta_1}$ (95\% CI) & $\hat{\beta_2}$ (95\% CI) & $\hat{\theta_0}$ (95\% CI) & $\hat{\theta_1}$ (95\% CI) & $\hat{\theta_2}$ (95\% CI)\\
\hline
\multirow{27}{*}{0.5} & \multirow{9}{*}{2} & \multirow{3}{*}{-0.5} & -2 & 0.5 (0.2, 0.9) & 2 (1.6, 2.5)   & 3 (2.5, 3.7)   & -0.5 (-0.8, -0.2) & -2 (-2.5, -1.7)   & -3.1 (-3.6, -2.6) \\
                      &                    &                       & -3 & 0.5 (0.2, 0.9) & 2 (1.6, 2.5)   & 3 (2.5, 3.7)   & -0.5 (-0.8, -0.2) & -3 (-3.6, -2.5)   & -3 (-3.6, -2.5)   \\
                      &                    &                       & -4 & 0.5 (0.2, 0.8) & 2 (1.6, 2.5)   & 3 (2.5, 3.6)   & -0.5 (-0.8, -0.2) & -3.9 (-4.7, -3.3) & -2.9 (-3.5, -2.4) \\
                      \cline{3-10}
                      &                    & \multirow{3}{*}{-1}   & -2 & 0.5 (0.2, 0.9) & 2 (1.6, 2.5)   & 3 (2.4, 3.7)   & -1 (-1.3, -0.7)   & -2 (-2.5, -1.7)   & -3.1 (-3.6, -2.6) \\
                      &                    &                       & -3 & 0.5 (0.2, 0.9) & 2 (1.6, 2.5)   & 3 (2.4, 3.7)   & -1 (-1.3, -0.7)   & -3 (-3.6, -2.5)   & -3 (-3.6, -2.5)   \\
                      &                    &                       & -4 & 0.5 (0.2, 0.9) & 2 (1.6, 2.5)   & 3.1 (2.5, 3.7) & -1 (-1.3, -0.7)   & -3.8 (-4.6, -3.2) & -2.9 (-3.5, -2.4) \\
                      \cline{3-10}
                      &                    & \multirow{3}{*}{-2}   & -2 & 0.5 (0.1, 1)   & 2 (1.5, 2.6)   & 3 (2.3, 3.8)   & -2 (-2.4, -1.7)   & -2 (-2.5, -1.6)   & -3.1 (-3.6, -2.6) \\
                      &                    &                       & -3 & 0.5 (0.1, 0.9) & 2 (1.5, 2.6)   & 3 (2.4, 3.8)   & -2 (-2.3, -1.6)   & -3 (-3.6, -2.4)   & -3 (-3.6, -2.5)   \\
                      &                    &                       & -4 & 0.5 (0.2, 0.9) & 2 (1.5, 2.5)   & 3 (2.4, 3.7)   & -1.9 (-2.3, -1.6) & -3.9 (-4.7, -3.3) & -2.9 (-3.5, -2.4) \\
                      \cline{2-10}
                      & \multirow{9}{*}{3} & \multirow{3}{*}{-0.5} & -2 & 0.5 (0.2, 0.9) & 3 (2.4, 3.7)   & 3 (2.4, 3.7)   & -0.5 (-0.8, -0.2) & -2 (-2.4, -1.7)   & -3 (-3.6, -2.6)   \\
                      &                    &                       & -3 & 0.5 (0.2, 0.9) & 3 (2.5, 3.7)   & 3 (2.4, 3.6)   & -0.5 (-0.8, -0.2) & -3 (-3.6, -2.5)   & -3 (-3.6, -2.5)   \\
                      &                    &                       & -4 & 0.5 (0.2, 0.8) & 3 (2.5, 3.6)   & 3 (2.4, 3.6)   & -0.5 (-0.8, -0.2) & -3.9 (-4.7, -3.3) & -2.9 (-3.5, -2.4) \\
                      \cline{3-10}
                      &                    & \multirow{3}{*}{-1}   & -2 & 0.5 (0.1, 0.9) & 2.9 (2.3, 3.7) & 2.9 (2.3, 3.7) & -1 (-1.2, -0.7)   & -2 (-2.4, -1.6)   & -3 (-3.5, -2.5)   \\
                      &                    &                       & -3 & 0.5 (0.2, 0.9) & 2.9 (2.4, 3.6) & 3 (2.4, 3.6)   & -1 (-1.3, -0.7)   & -3 (-3.6, -2.5)   & -3 (-3.6, -2.5)   \\
                      &                    &                       & -4 & 0.5 (0.2, 0.8) & 3 (2.4, 3.6)   & 3 (2.4, 3.6)   & -1 (-1.3, -0.7)   & -3.9 (-4.7, -3.3) & -2.9 (-3.5, -2.4) \\
                      \cline{3-10}
                      &                    & \multirow{3}{*}{-2}   & -2 & 0.4 (0, 0.9)   & 2.9 (2.2, 3.7) & 2.9 (2.2, 3.7) & -2 (-2.4, -1.7)   & -2 (-2.4, -1.6)   & -3 (-3.6, -2.5)   \\
                      &                    &                       & -3 & 0.5 (0.1, 1)   & 3 (2.3, 3.8)   & 3 (2.3, 3.8)   & -2 (-2.4, -1.6)   & -3 (-3.6, -2.5)   & -3 (-3.6, -2.5)   \\
                      &                    &                       & -4 & 0.5 (0.1, 0.9) & 3 (2.4, 3.7)   & 3 (2.4, 3.7)   & -2 (-2.4, -1.6)   & -3.9 (-4.7, -3.3) & -2.9 (-3.5, -2.4) \\
                      \cline{2-10}
                      & \multirow{9}{*}{4} & \multirow{3}{*}{-0.5} & -2 & 0.4 (0.1, 0.8) & 3.9 (3.2, 4.8) & 2.9 (2.4, 3.6) & -0.5 (-0.7, -0.2) & -2 (-2.4, -1.6)   & -3 (-3.5, -2.5)   \\
                      &                    &                       & -3 & 0.5 (0.1, 0.8) & 3.9 (3.2, 4.7) & 2.9 (2.3, 3.6) & -0.5 (-0.8, -0.2) & -3 (-3.5, -2.5)   & -3 (-3.5, -2.5)   \\
                      &                    &                       & -4 & 0.5 (0.2, 0.9) & 4 (3.2, 4.8)   & 3 (2.4, 3.6)   & -0.5 (-0.8, -0.2) & -3.9 (-4.6, -3.3) & -2.9 (-3.5, -2.4) \\
                      \cline{3-10}
                      &                    & \multirow{3}{*}{-1}   & -2 & 0.5 (0.1, 0.9) & 3.8 (3.1, 4.8) & 2.9 (2.3, 3.6) & -1 (-1.3, -0.7)   & -2 (-2.4, -1.6)   & -3 (-3.5, -2.5)   \\
                      &                    &                       & -3 & 0.5 (0.1, 0.9) & 3.9 (3.1, 4.8) & 2.9 (2.3, 3.6) & -1 (-1.3, -0.7)   & -3 (-3.5, -2.5)   & -3 (-3.5, -2.5)   \\
                      &                    &                       & -4 & 0.5 (0.1, 0.9) & 3.9 (3.2, 4.8) & 2.9 (2.4, 3.6) & -1 (-1.3, -0.7)   & -4 (-4.7, -3.3)   & -3 (-3.6, -2.5)   \\
                      \cline{3-10}
                      &                    & \multirow{3}{*}{-2}   & -2 & 0.4 (0, 0.9)   & 3.8 (2.9, 4.9) & 2.9 (2.2, 3.7) & -2 (-2.4, -1.7)   & -2 (-2.5, -1.7)   & -3 (-3.6, -2.5)   \\
                      &                    &                       & -3 & 0.5 (0.1, 0.9) & 3.9 (3.1, 5)   & 2.9 (2.3, 3.7) & -2 (-2.4, -1.6)   & -3 (-3.6, -2.5)   & -3 (-3.6, -2.5)   \\
                      &                    &                       & -4 & 0.5 (0.1, 0.9) & 3.9 (3.1, 4.8) & 2.9 (2.3, 3.7) & -2 (-2.4, -1.6)   & -3.9 (-4.7, -3.3) & -2.9 (-3.5, -2.4) \\
\hline
\multirow{27}{*}{1} & \multirow{9}{*}{2} & \multirow{3}{*}{-0.5}   & -2 & 1 (0.7, 1.5)   & 2 (1.6, 2.5)   & 3 (2.5, 3.7)   & -0.5 (-0.8, -0.3) & -2 (-2.4, -1.7)   & -3 (-3.6, -2.6)   \\
                      &                    &                       & -3 & 1 (0.7, 1.4)   & 2 (1.6, 2.4)   & 3 (2.5, 3.6)   & -0.5 (-0.7, -0.2) & -3 (-3.6, -2.5)   & -3 (-3.5, -2.5)   \\
                      &                    &                       & -4 & 1 (0.7, 1.4)   & 2 (1.6, 2.5)   & 3 (2.5, 3.6)   & -0.5 (-0.8, -0.2) & -3.9 (-4.6, -3.3) & -2.9 (-3.5, -2.4) \\
                      \cline{3-10}
                      &                    & \multirow{3}{*}{-1}   & -2 & 1 (0.6, 1.4)   & 2 (1.5, 2.5)   & 3 (2.4, 3.7)   & -1 (-1.2, -0.7)   & -2 (-2.4, -1.7)   & -3 (-3.6, -2.6)   \\
                      &                    &                       & -3 & 1 (0.6, 1.4)   & 2 (1.6, 2.5)   & 3 (2.4, 3.6)   & -1 (-1.3, -0.7)   & -3 (-3.6, -2.5)   & -3 (-3.6, -2.5)   \\
                      &                    &                       & -4 & 1 (0.7, 1.4)   & 2 (1.6, 2.5)   & 3 (2.5, 3.7)   & -1 (-1.3, -0.7)   & -3.9 (-4.6, -3.3) & -2.9 (-3.5, -2.4) \\
                      \cline{3-10}
                      &                    & \multirow{3}{*}{-2}   & -2 & 1 (0.5, 1.6)   & 2 (1.4, 2.6)   & 3 (2.3, 3.8)   & -2 (-2.3, -1.7)   & -2 (-2.4, -1.6)   & -3 (-3.6, -2.5)   \\
                      &                    &                       & -3 & 1 (0.6, 1.5)   & 2 (1.5, 2.6)   & 3 (2.4, 3.8)   & -2 (-2.4, -1.7)   & -3 (-3.5, -2.5)   & -3 (-3.6, -2.5)   \\
                      &                    &                       & -4 & 1 (0.6, 1.5)   & 2 (1.5, 2.5)   & 3 (2.4, 3.7)   & -1.9 (-2.3, -1.6) & -3.9 (-4.6, -3.3) & -2.9 (-3.5, -2.4) \\
                      \cline{2-10}
                      & \multirow{9}{*}{3} & \multirow{3}{*}{-0.5} & -2 & 1 (0.6, 1.4)   & 2.9 (2.4, 3.6) & 2.9 (2.3, 3.6) & -0.5 (-0.7, -0.2) & -2 (-2.4, -1.7)   & -3.1 (-3.6, -2.6) \\
                      &                    &                       & -3 & 1 (0.6, 1.4)   & 3 (2.4, 3.6)   & 3 (2.4, 3.6)   & -0.5 (-0.8, -0.2) & -3 (-3.5, -2.5)   & -3 (-3.5, -2.5)   \\
                      &                    &                       & -4 & 1 (0.7, 1.4)   & 3 (2.5, 3.6)   & 3 (2.4, 3.6)   & -0.5 (-0.8, -0.2) & -4 (-4.7, -3.4)   & -3 (-3.6, -2.5)   \\
                      \cline{3-10}
                      &                    & \multirow{3}{*}{-1}   & -2 & 0.9 (0.5, 1.4) & 2.9 (2.3, 3.6) & 3 (2.4, 3.7)   & -1 (-1.2, -0.7)   & -2 (-2.4, -1.7)   & -3 (-3.6, -2.6)   \\
                      &                    &                       & -3 & 1 (0.6, 1.4)   & 3 (2.4, 3.7)   & 3 (2.4, 3.7)   & -1 (-1.3, -0.7)   & -3 (-3.6, -2.5)   & -3 (-3.5, -2.5)   \\
                      &                    &                       & -4 & 1 (0.6, 1.4)   & 3 (2.4, 3.6)   & 3 (2.4, 3.6)   & -1 (-1.3, -0.7)   & -3.9 (-4.6, -3.3) & -2.9 (-3.5, -2.4) \\
                      \cline{3-10}
                      &                    & \multirow{3}{*}{-2}   & -2 & 1 (0.5, 1.6)   & 3 (2.3, 3.9)   & 2.9 (2.3, 3.8) & -2 (-2.4, -1.7)   & -2 (-2.4, -1.7)   & -3 (-3.6, -2.6)   \\
                      &                    &                       & -3 & 0.9 (0.5, 1.5) & 2.9 (2.3, 3.7) & 2.9 (2.3, 3.7) & -2 (-2.3, -1.6)   & -3 (-3.6, -2.5)   & -3 (-3.6, -2.5)   \\
                      &                    &                       & -4 & 0.9 (0.5, 1.4) & 2.9 (2.3, 3.7) & 2.9 (2.3, 3.7) & -1.9 (-2.3, -1.6) & -3.9 (-4.7, -3.3) & -2.9 (-3.5, -2.4) \\
                      \cline{2-10}
                      & \multirow{9}{*}{4} & \multirow{3}{*}{-0.5} & -2 & 0.9 (0.6, 1.4) & 3.9 (3.1, 4.7) & 2.9 (2.3, 3.6) & -0.5 (-0.7, -0.3) & -2 (-2.4, -1.7)   & -3 (-3.5, -2.6)   \\
                      &                    &                       & -3 & 1 (0.6, 1.4)   & 3.9 (3.2, 4.8) & 2.9 (2.4, 3.6) & -0.5 (-0.8, -0.2) & -3 (-3.6, -2.5)   & -3 (-3.6, -2.6)   \\
                      &                    &                       & -4 & 0.9 (0.6, 1.3) & 3.9 (3.2, 4.7) & 2.9 (2.4, 3.6) & -0.5 (-0.7, -0.2) & -4 (-4.7, -3.4)   & -3 (-3.5, -2.5)   \\
                      \cline{3-10}
                      &                    & \multirow{3}{*}{-1}   & -2 & 0.9 (0.5, 1.4) & 3.9 (3.1, 4.8) & 2.8 (2.2, 3.6) & -1 (-1.3, -0.7)   & -2 (-2.4, -1.7)   & -3 (-3.6, -2.6)   \\
                      &                    &                       & -3 & 1 (0.6, 1.4)   & 3.9 (3.2, 4.8) & 2.9 (2.4, 3.7) & -1 (-1.2, -0.7)   & -3 (-3.6, -2.5)   & -3 (-3.6, -2.5)   \\
                      &                    &                       & -4 & 1 (0.6, 1.4)   & 3.9 (3.2, 4.7) & 2.9 (2.3, 3.6) & -0.9 (-1.2, -0.7) & -4 (-4.7, -3.3)   & -3 (-3.5, -2.5)   \\
                      \cline{3-10}
                      &                    & \multirow{3}{*}{-2}   & -2 & 0.9 (0.4, 1.5) & 3.8 (2.9, 4.9) & 2.9 (2.2, 3.7) & -2 (-2.3, -1.7)   & -2 (-2.4, -1.7)   & -3 (-3.5, -2.5)   \\
                      &                    &                       & -3 & 1 (0.5, 1.5)   & 3.9 (3, 4.9)   & 2.9 (2.2, 3.7) & -2 (-2.4, -1.7)   & -3 (-3.5, -2.5)   & -3 (-3.6, -2.5)   \\
                      &                    &                       & -4 & 0.9 (0.5, 1.4) & 3.9 (3.1, 4.9) & 2.9 (2.3, 3.7) & -1.9 (-2.3, -1.6) & -3.9 (-4.6, -3.3) & -3 (-3.5, -2.5)   \\
\hline
\end{tabular}
\end{table*}

\begin{table*}[h!]
\tiny\sf\centering
\caption{Simulation study results including covariates (II).\label{tab3_2}}
\begin{tabular}{p{0.4cm}p{0.4cm}p{0.4cm}p{0.4cm}p{1.50cm}p{1.50cm}p{1.50cm}p{1.50cm}p{1.50cm}p{1.50cm}} % columns
\hline                                          
$\beta_0$ & $\beta_1$ & $\theta_0$ & $\theta_1$ & $\hat{\beta_0}$ (95\% CI) & $\hat{\beta_1}$ (95\% CI) & $\hat{\beta_2}$ (95\% CI) & $\hat{\theta_0}$ (95\% CI) & $\hat{\theta_1}$ (95\% CI) & $\hat{\theta_2}$ (95\% CI)\\
\hline
\multirow{27}{*}{2} & \multirow{9}{*}{2} & \multirow{3}{*}{-0.5}   & -2 & 1.9 (1.5, 2.5) & 2 (1.5, 2.5)   & 3 (2.4, 3.7)   & -0.5 (-0.7, -0.2) & -2 (-2.4, -1.7)   & -3.1 (-3.6, -2.6) \\
                      &                    &                       & -3 & 2 (1.6, 2.6)   & 2 (1.6, 2.5)   & 3 (2.4, 3.6)   & -0.5 (-0.8, -0.3) & -3 (-3.6, -2.6)   & -3 (-3.5, -2.6)   \\
                      &                    &                       & -4 & 2 (1.6, 2.5)   & 2 (1.6, 2.5)   & 3 (2.4, 3.6)   & -0.5 (-0.8, -0.2) & -3.9 (-4.6, -3.4) & -3 (-3.5, -2.5)   \\
                      \cline{3-10}
                      &                    & \multirow{3}{*}{-1}   & -2 & 1.9 (1.4, 2.5) & 1.9 (1.5, 2.5) & 2.9 (2.3, 3.6) & -1 (-1.2, -0.8)   & -2 (-2.4, -1.7)   & -3.1 (-3.6, -2.6) \\
                      &                    &                       & -3 & 2 (1.5, 2.6)   & 1.9 (1.5, 2.5) & 2.9 (2.4, 3.6) & -1 (-1.2, -0.7)   & -3 (-3.5, -2.6)   & -3 (-3.5, -2.5)   \\
                      &                    &                       & -4 & 2 (1.6, 2.6)   & 2 (1.6, 2.5)   & 3 (2.4, 3.7)   & -1 (-1.3, -0.7)   & -4 (-4.7, -3.4)   & -3 (-3.5, -2.5)   \\
                      \cline{3-10}
                      &                    & \multirow{3}{*}{-2}   & -2 & 1.9 (1.3, 2.7) & 2 (1.5, 2.7)   & 2.9 (2.2, 3.8) & -2 (-2.3, -1.7)   & -2 (-2.4, -1.7)   & -3 (-3.5, -2.6)   \\
                      &                    &                       & -3 & 2 (1.4, 2.7)   & 2 (1.5, 2.6)   & 2.9 (2.3, 3.8) & -2 (-2.3, -1.7)   & -3 (-3.5, -2.5)   & -3 (-3.5, -2.5)   \\
                      &                    &                       & -4 & 1.9 (1.4, 2.6) & 1.9 (1.5, 2.5) & 2.9 (2.3, 3.7) & -2 (-2.3, -1.7)   & -4 (-4.7, -3.4)   & -3 (-3.5, -2.5)   \\
                      \cline{2-10}
                      & \multirow{9}{*}{3} & \multirow{3}{*}{-0.5} & -2 & 2 (1.5, 2.6)   & 3 (2.4, 3.7)   & 3 (2.4, 3.7)   & -0.5 (-0.7, -0.2) & -2 (-2.4, -1.7)   & -3 (-3.5, -2.6)   \\
                      &                    &                       & -3 & 2 (1.5, 2.5)   & 3 (2.4, 3.6)   & 2.9 (2.4, 3.6) & -0.5 (-0.7, -0.3) & -3 (-3.5, -2.6)   & -3 (-3.5, -2.5)   \\
                      &                    &                       & -4 & 1.9 (1.5, 2.5) & 3 (2.4, 3.6)   & 2.9 (2.4, 3.6) & -0.5 (-0.7, -0.2) & -3.9 (-4.6, -3.4) & -2.9 (-3.5, -2.5) \\
                      \cline{3-10}
                      &                    & \multirow{3}{*}{-1}   & -2 & 1.9 (1.4, 2.6) & 2.9 (2.3, 3.7) & 2.9 (2.3, 3.7) & -1 (-1.3, -0.8)   & -2 (-2.4, -1.7)   & -3 (-3.5, -2.6)   \\
                      &                    &                       & -3 & 1.9 (1.4, 2.5) & 2.9 (2.3, 3.6) & 2.9 (2.3, 3.6) & -1 (-1.3, -0.7)   & -3 (-3.5, -2.5)   & -3 (-3.5, -2.6)   \\
                      &                    &                       & -4 & 1.9 (1.5, 2.5) & 2.9 (2.3, 3.6) & 2.9 (2.4, 3.6) & -1 (-1.3, -0.7)   & -4 (-4.7, -3.4)   & -3 (-3.6, -2.5)   \\
                      \cline{3-10}
                      &                    & \multirow{3}{*}{-2}   & -2 & 1.9 (1.3, 2.6) & 2.9 (2.2, 3.7) & 2.9 (2.2, 3.8) & -2 (-2.3, -1.7)   & -2 (-2.4, -1.7)   & -3 (-3.5, -2.6)   \\
                      &                    &                       & -3 & 1.9 (1.3, 2.6) & 2.9 (2.3, 3.7) & 2.9 (2.2, 3.7) & -2 (-2.3, -1.7)   & -3 (-3.5, -2.6)   & -3 (-3.5, -2.5)   \\
                      &                    &                       & -4 & 2 (1.4, 2.6)   & 2.9 (2.3, 3.7) & 2.9 (2.3, 3.7) & -1.9 (-2.3, -1.6) & -3.9 (-4.5, -3.3) & -2.9 (-3.4, -2.5) \\
                      \cline{2-10}
                      & \multirow{9}{*}{4} & \multirow{3}{*}{-0.5} & -2 & 1.9 (1.4, 2.5) & 3.8 (3.1, 4.7) & 2.9 (2.3, 3.6) & -0.5 (-0.7, -0.3) & -2 (-2.4, -1.7)   & -3.1 (-3.5, -2.6) \\
                      &                    &                       & -3 & 1.9 (1.4, 2.5) & 3.8 (3.1, 4.7) & 2.9 (2.3, 3.6) & -0.5 (-0.7, -0.2) & -3 (-3.6, -2.6)   & -3 (-3.6, -2.6)   \\
                      &                    &                       & -4 & 1.9 (1.4, 2.4) & 3.8 (3.1, 4.7) & 2.9 (2.3, 3.6) & -0.5 (-0.7, -0.2) & -4 (-4.7, -3.4)   & -3 (-3.5, -2.5)   \\
                      \cline{3-10}
                      &                    & \multirow{3}{*}{-1}   & -2 & 1.9 (1.4, 2.6) & 3.8 (3, 4.8)   & 2.9 (2.3, 3.7) & -1 (-1.2, -0.8)   & -2 (-2.4, -1.7)   & -3 (-3.5, -2.6)   \\
                      &                    &                       & -3 & 1.9 (1.4, 2.5) & 3.9 (3.1, 4.8) & 2.9 (2.3, 3.6) & -1 (-1.3, -0.7)   & -3 (-3.5, -2.6)   & -3 (-3.5, -2.6)   \\
                      &                    &                       & -4 & 2 (1.5, 2.5)   & 3.9 (3.2, 4.8) & 2.9 (2.3, 3.6) & -1 (-1.2, -0.7)   & -3.9 (-4.6, -3.4) & -3 (-3.5, -2.5)   \\
                      \cline{3-10}
                      &                    & \multirow{3}{*}{-2}   & -2 & 1.8 (1.2, 2.6) & 3.8 (2.9, 4.9) & 2.8 (2.1, 3.7) & -2 (-2.3, -1.7)   & -2 (-2.4, -1.7)   & -3 (-3.5, -2.6)   \\
                      &                    &                       & -3 & 1.8 (1.3, 2.5) & 3.7 (2.9, 4.7) & 2.8 (2.1, 3.6) & -2 (-2.3, -1.7)   & -3.1 (-3.6, -2.6) & -3 (-3.6, -2.6)   \\
                      &                    &                       & -4 & 1.9 (1.4, 2.6) & 3.8 (3, 4.8)   & 2.9 (2.2, 3.7) & -1.9 (-2.3, -1.6) & -3.9 (-4.6, -3.3) & -2.9 (-3.5, -2.5) \\
\hline
\end{tabular}
\end{table*}

The \textit{R} code used for the simulation is available as Supplementary Material.

\section{Discussion}\label{discussion}
The proposed methodology is able to distinguish two different sources of zeros (structural and non-structural) from dichotomous data in a Bayesian framework by assuming priors with different parameters on proportion of structural and non-structural zeros. Furthermore, since it is freely available as an \textit{R} package, it is easily usable for any researcher who needs to adjust this type of data and easily modifiable for more advanced users who need to adapt the model to their context, for example with different choices of the \textit{prior} distributions of $\omega$ and $p$.

The approach used to analyze the SP is an important topic. Some studies include all working population to estimate SP, whilst other exclude ``healthy'' workers. As result, different conclusions in terms of prevalence and associated factors are obtained \cite{Navarro2019}. SP is an outcome resulting from mixing two phenomena, i.e. health status and exercise of rights. Health status plays a role regarding the fact of being exposed; and, among the exposed, the lack of the exercise of the right to take a sick leave determines SP. Using the proposed ZIB approach one could describe, in a single analysis, both phenomena: first, which factors are associated to the exposure to presenteeism (to be ``sick'', factors related to health status), and after that, which factors increase the probability of being presenteeist among the exposed workers.

The simulation study shows that, even with relatively small sample sizes the model is capable of producing reasonable estimates for the parameters involved in both the zero inflated and non zero inflated processes. As expected, the credibility intervals length diminishes with sample size while their coverage grows.

%\begin{acks}
%David Mori\~na acknowledges financial support from the Spanish Ministry of Economy and Competitiveness, through the Mar\'ia de Maeztu Programme for Units %of Excellence in R\&D (MDM-2014-0445) and from Fundaci\'on Santander Universidades.
%\end{acks}

\bibliographystyle{plain}
\bibliography{morina_puig_navarro}

\begin{thebibliography}{1}

\bibitem{Barbu2015}
Adrian Barbu, Tianfu Wu, and Ying~Nian Wu.
\newblock {Learning mixtures of bernoulli templates by two-round EM with
  performance guarantee}.
\newblock {\em Electronic Journal of Statistics}, 8:3004--3030, 2015.

\bibitem{Diop2016}
Aba Diop, Aliou Diop, and Jean~Fran{\c{c}}ois Dupuy.
\newblock {Simulation-based Inference in a Zero-inflated Bernoulli Regression
  Model}.
\newblock {\em Communications in Statistics: Simulation and Computation},
  45(10):3597--3614, 2016.

\bibitem{Johns2010}
Gary Johns.
\newblock {Presenteeism in the workplace: A review and research agenda}.
\newblock {\em Journal of Organizational Behavior}, 31(4):519--542, may 2010.

\bibitem{Navarro2019}
A.~Navarro, S.~Salas-Nic{\'{a}}s, C.~Llorens, S.~Moncada,
  E.~Molinero-Ru{\'{i}}z, and D.~Mori{\~{n}}a.
\newblock {Sickness presenteeism: Are we sure about what we are studying? A
  research based on a literature review and an empirical illustration}.
\newblock {\em American Journal of Industrial Medicine}, 62(7), 2019.

\bibitem{Navarro2018}
Albert Navarro, Sergio Salas-Nic{\'{a}}s, Salvador Moncada, Clara Llorens, and
  Emilia Molinero-Ruiz.
\newblock {Prevalence, associated factors and reasons for sickness
  presenteeism: a cross-sectional nationally representative study of salaried
  workers in Spain, 2016.}
\newblock {\em BMJ open}, 8(7):e021212, 2018.

\bibitem{cran}
{R Core Team}.
\newblock {\em R: A Language and Environment for Statistical Computing}.
\newblock R Foundation for Statistical Computing, Vienna, Austria, 2021.

\bibitem{Sun2007}
Zhuoxin Sun, Ori Rosen, and Allan~R. Sampson.
\newblock {Multivariate Bernoulli mixture models with application to postmortem
  tissue studies in schizophrenia}.
\newblock {\em Biometrics}, 63(3):901--909, sep 2007.

\bibitem{Vives2013}
Alejandra Vives, Marcelo Amable, Montserrat Ferrer, Salvador Moncada, Clara
  Llorens, Carles Muntaner, Fernando~G Benavides, and Joan Benach.
\newblock {Employment precariousness and poor mental health: evidence from
  Spain on a new social determinant of health.}
\newblock {\em Journal of environmental and public health}, 2013:978656, jan
  2013.

\bibitem{pscl}
Achim Zeileis, Christian Kleiber, and Simon Jackman.
\newblock Regression models for count data in {R}.
\newblock {\em Journal of Statistical Software}, 27(8), 2008.

\end{thebibliography}

\section*{Appendix A}
A second simulation was conducted to evaluate the performance of the proposed methodology when there are no covariates involved. In this case, similarly to the simulation process described in the previous section \textit{Simulation study}, 100 random samples were generated for each considered sample size ($n = 500, 1500$), proportion of structural zeros ($1-\omega = 0.6, 0.7, 0.8, 0.9$) and proportion of non structural zeros ($1-p = 0.1, 0.2, 0.3, 0.4$). For each random sample, the marginal posterior distributions of the parameters have been summarised by their median and percentiles 2.5\% and 97.5\%.

Table~\ref{tab4} shows, for each combination of parameters, the average estimates and upper and lower limits of the 95\% credibility intervals. 

\begin{table}[h]
\small\sf\centering
\caption{Simulation study results with no covariates.\label{tab4}}
\begin{tabular}{lllll}
\hline
$n$                  & $\omega$             & $p$ & $\hat{\omega}$ (95\% CI) & $\hat{p}$ (95\% CI) \\
\hline
\multirow{16}{*}{500}  & \multirow{4}{*}{0.1} & 0.6 & 0.09 (0.06, 0.15) & 0.7 (0.51, 0.99)  \\
                       &                      & 0.7 & 0.09 (0.05, 0.14) & 0.71 (0.51, 0.98) \\
                       &                      & 0.8 & 0.13 (0.08, 0.2)  & 0.72 (0.51, 0.98) \\
                       &                      & 0.9 & 0.13 (0.08, 0.21) & 0.71 (0.51, 0.98) \\
                       \cline{2-5}                                          
                       & \multirow{4}{*}{0.2} & 0.6 & 0.16 (0.1, 0.24)  & 0.71 (0.51, 0.98) \\
                       &                      & 0.7 & 0.21 (0.14, 0.31) & 0.7 (0.51, 0.98)  \\
                       &                      & 0.8 & 0.23 (0.15, 0.34) & 0.7 (0.51, 0.98)  \\
                       &                      & 0.9 & 0.26 (0.17, 0.39) & 0.7 (0.51, 0.98)  \\
                       \cline{2-5}                                          
                       & \multirow{4}{*}{0.3} & 0.6 & 0.24 (0.16, 0.35) & 0.71 (0.51, 0.98) \\
                       &                      & 0.7 & 0.27 (0.18, 0.39) & 0.71 (0.51, 0.98) \\
                       &                      & 0.8 & 0.37 (0.27, 0.49) & 0.76 (0.55, 0.99) \\
                       &                      & 0.9 & 0.34 (0.24, 0.48) & 0.73 (0.51, 0.98) \\
                       \cline{2-5}                                          
                       & \multirow{4}{*}{0.4} & 0.6 & 0.33 (0.23, 0.48) & 0.71 (0.51, 0.98) \\
                       &                      & 0.7 & 0.38 (0.27, 0.49) & 0.75 (0.54, 0.99) \\
                       &                      & 0.8 & 0.39 (0.3, 0.49)  & 0.78 (0.6, 0.99)  \\
                       &                      & 0.9 & 0.45 (0.38, 0.5)  & 0.9 (0.77, 1)     \\
                       \cline{1-5}                                          
\multirow{16}{*}{1500} & \multirow{4}{*}{0.1} & 0.6 & 0.1 (0.07, 0.15)  & 0.7 (0.51, 0.98)  \\
                       &                      & 0.7 & 0.09 (0.06, 0.14) & 0.71 (0.51, 0.98) \\
                       &                      & 0.8 & 0.1 (0.07, 0.16)  & 0.7 (0.51, 0.99)  \\
                       &                      & 0.9 & 0.13 (0.09, 0.2)  & 0.71 (0.51, 0.98) \\
                       \cline{2-5}                                          
                       & \multirow{4}{*}{0.2} & 0.6 & 0.14 (0.1, 0.21)  & 0.71 (0.51, 0.98) \\
                       &                      & 0.7 & 0.22 (0.15, 0.32) & 0.71 (0.51, 0.98) \\
                       &                      & 0.8 & 0.23 (0.16, 0.34) & 0.71 (0.51, 0.98) \\
                       &                      & 0.9 & 0.24 (0.16, 0.34) & 0.71 (0.51, 0.98) \\
                       \cline{2-5}                                          
                       & \multirow{4}{*}{0.3} & 0.6 & 0.25 (0.17, 0.36) & 0.7 (0.51, 0.98)  \\
                       &                      & 0.7 & 0.29 (0.2, 0.42)  & 0.71 (0.51, 0.98) \\
                       &                      & 0.8 & 0.36 (0.26, 0.49) & 0.72 (0.52, 0.98) \\
                       &                      & 0.9 & 0.36 (0.27, 0.49) & 0.74 (0.54, 0.98) \\
                       \cline{2-5}                                          
                       & \multirow{4}{*}{0.4} & 0.6 & 0.35 (0.25, 0.49) & 0.71 (0.51, 0.99) \\
                       &                      & 0.7 & 0.38 (0.3, 0.49)  & 0.79 (0.61, 0.99) \\
                       &                      & 0.8 & 0.41 (0.32, 0.49) & 0.82 (0.65, 0.99) \\
                       &                      & 0.9 & 0.44 (0.38, 0.5)  & 0.88 (0.74, 0.99) \\
\hline
\end{tabular}
\end{table}

\end{document}